\documentstyle[10pt,draft,psfig]{nature}

\def\duck{G5.27--0.90}
\def\HI{H\,{\sc i}}
\def\kms{km~s$^{-1}$}
\def\la{\ifmmode\stackrel{<}{_{\sim}}\else$\stackrel{<}{_{\sim}}$\fi} 
\def\ga{\ifmmode\stackrel{>}{_{\sim}}\else$\stackrel{>}{_{\sim}}$\fi} 

\title{\LARGE \bf A large age for the pulsar B1757--24 from
an upper limit on its proper motion}
\author{B. M. Gaensler\affiliation[1]{Center for Space Research,
Massachusetts
Institute of Technology, Cambridge, MA 02139, USA}\affiliation[3]{Hubble
Fellow} and
D. A. Frail\affiliation[2]{National Radio Astronomy Observatory, P.O. Box
0, Socorro, NM 87801, USA}}

\dates{\date}{}
\headertitle{A large age for PSR B1757--24}
\mainauthor{Gaensler \& Frail}

\begin{document}

%

\normalsize

\summary{The ``characteristic age''\cite{mt77} of a pulsar usually is
considered to approximate its true age, but this assumption has led to
some puzzling results, including the fact that many pulsars with small
characteristic ages have no associated supernova
remnants\cite{bgl89,sgj99}.  The pulsar B1757--24 is located just beyond
the edge of a supernova remnant\cite{mdt85,mkj+91,fk91}; the properties
of the system indicate that the pulsar was born at the centre of the
remnant, but that it has subsequently overtaken the expanding
blast-wave\cite{mkj+91,fk91,sfs89,fkw94}.  With a characteristic age of
16,000 yr, this implies an expected\cite{fkw94} proper motion by the
pulsar of 63--80 milliarcseconds (mas) per year.  Here we report
observations of the nebula surrounding the pulsar which limit its
proper motion to less than 25~mas~yr$^{-1}$, implying a minimum age of
39,000 yr.  A more detailed analysis argues for a true age as great as
170,000 yr, significantly larger than the characteristic age. From this
result and other discrepancies associated with pulsars, we conclude
that characteristic ages seriously underestimate the true ages of
pulsars.}

\maketitle

\begin{figure}
\centerline{\psfig{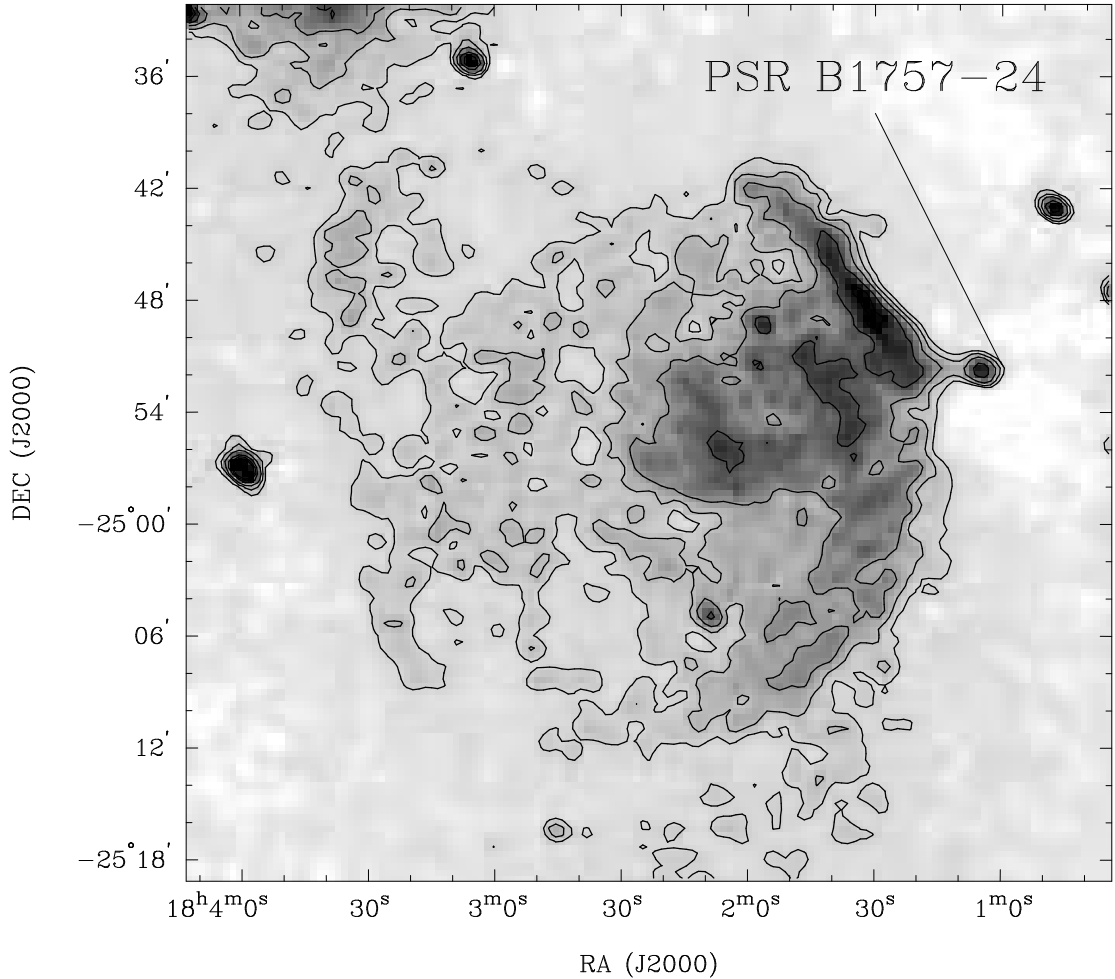}}
\caption{Radio emission from the supernova remnant G5.4--1.2.
The image corresponds to archival VLA data at an
observing wavelength of 90~cm;
contours are at levels of 10, 25, 50, 100, 150 and 200~mJy~beam$^{-1}$,
and the peak intensity is 150~mJy~beam$^{-1}$. The resolution
of the image is $60''\times45''$. The pulsar (whose position
is marked) is at the head of the compact nebula \duck\ to the west
of the supernova remnant.}
\label{fig_snr}
\end{figure}

The radio supernova remnant G5.4--1.2 is shown in Fig~\ref{fig_snr}: it
has an approximately circular radio morphology of diameter $\sim35'$,
but is much brighter and has a significantly flatter spectral index
along its western edge.\cite{fk91,fkw94,bh85,ckk+87} Approximately
$5'$ beyond this bright western rim is the 125-ms pulsar
B1757--24\cite{mdt85,mkj+91,fk91}, whose rotational frequency, $\nu$,
and frequency derivative, $\dot{\nu}$, can be used to derive a
characteristic age, $\tau \equiv \nu/2\dot{\nu} \approx 16$~kyr.
Provided that a pulsar's initial frequency is much larger
than $\nu$, that the pulsar slows down by magnetic dipole braking, and
that its magnetic dipole moment and moment of inertia do not evolve
with time, then $\tau \approx t_p$, where $t_p$ is the pulsar's
true age\cite{mt77}.

Surrounding the pulsar is the radio nebula \duck, whose properties
are consistent with it being powered by the pulsar's relativistic
wind\cite{fk91,fkw94}.  \duck\ has a cometary morphology with a tail
pointing back towards the centre of G5.4--1.2, an appearance which
strongly suggests that the pulsar is travelling away from the centre of
G5.4--1.2 at high velocity.  The characteristic age of B1757--24 and
its positional offset from the centre of G5.4--1.2 together imply a
proper motion of 63--80 milliarcsec~yr$^{-1}$ (ref.\ 8\nocite{fkw94}).
For a distance to the system of 5~kpc\cite{fkw94}, this corresponds
to a transverse velocity for the pulsar in the range 1500--1900~\kms,
significantly higher than typical pulsar velocities of $\sim$500~\kms\
(refs 11, 12\nocite{ll94,cc98}).

\begin{figure}
\centerline{\psfig{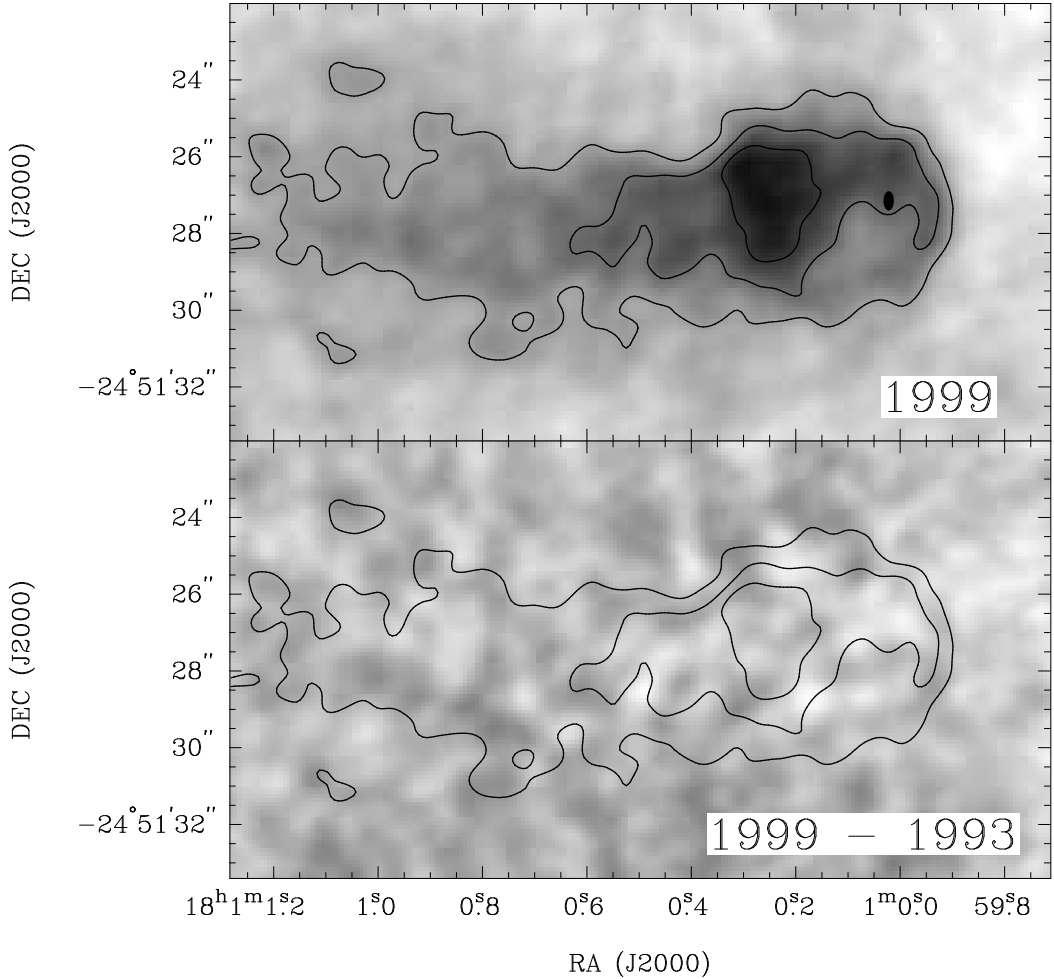}}
\caption{Radio emission from the western tip of the pulsar-powered
nebula \duck. The upper panel
represents the 1999 image of \duck, with corresponding contour levels
at 0.75, 1.5 and 2.25~mJy~beam$^{-1}$.  The ellipse near the head of
the nebula marks the position of PSR~B1757--24, (J2000) RA~18$^{\rm
h}$01$^{\rm m}$00.023(2)$^{\rm s}$, Dec~$-24^\circ51'27.15(5)''$, which
was determined by 20~cm A-array observations in Aug 1999 (uncertainties
in the position of the pulsar are a factor of five smaller than the extent
of the ellipse). The lower panel shows the difference between the 1999
and 1993 images, on which are superimposed the same contour levels as in
the upper panel.  
To compare the two epochs in a quantitative fashion, we took
the deconvolved model from each epoch, computed its two-dimensional
Fourier transform, and then sampled this transform with the transfer
function corresponding to the $u-v$ coverage of the 1993 data. Assuming
fidelity in the original deconvolution, this results in two data-sets
which have identical $u-v$ coverage and sensitivity, and differ only in
the actual brightness distribution between the two epochs.
We searched for proper motion between these two images
using the {\tt MIRIAD}\ task {\tt IMDIFF}\cite{sk98}, which finds
the shift in $x$ and $y$ which minimises\cite{key81} the rms noise of
the difference map between the two data-sets. Shifts corresponding to
a non-integral number of pixels were computed using cubic convolution
interpolation\cite{pow64}.  To estimate the uncertainties associated
with this approach, we repeated this comparison 40 times, each time
adding Gaussian noise of rms 30~$\mu$Jy~beam$^{-1}$ (corresponding to
the sensitivity of the Epoch~1 data) to one of the images.}
\label{fig_duck}
\end{figure}

To look for this expected proper motion, we have undertaken
multi-epoch observations of G5.27--0.90 with the Very Large Array
(VLA).  Our observations 
are summarised in Table~1; the two epochs are separated
by 6.7~yr.  The two data-sets were processed in identical fashion,
with results shown in Fig~\ref{fig_duck}.  If G5.27-0.90 is associated
with the pulsar, then the expected shift of the nebula between these
two epochs is 422--526~milliarcsec in a westwardly direction; it is
immediately apparent, even by eye, that motion of this magnitude is
not observed.  The two epochs can be compared
in a more quantitative fashion using the approach described
in the legend to Fig~\ref{fig_duck};
we find an offset between the two epochs (in the
sense [Epoch 2]--[Epoch 1]) of $\Delta$(RA)~$=+39\pm41$~milliarcsec and
$\Delta$(Dec)~$=-9\pm34$~milliarcsec, where contributions from both the
noise in the images and estimated errors in the phase calibration of
the data have been combined in quadrature.
We can thus rule out the proper motion which would be expected 
if $t_p = \tau$ 
at the 11--14~$\sigma$ level, independent of the distance
adopted to either the pulsar or the SNR.

We now consider various explanations for this surprising discrepancy,
and show that the most reasonable explanation is 
that $t_p \gg \tau$.
In the following discussion, we only consider proper motion in a westward
direction (as implied by the morphology of \duck), and adopt a
5-$\sigma$ upper limit on any shift between epochs of 166~milliarcsec,
corresponding to a maximum proper motion of
24.8~milliarcsec~yr$^{-1}$.  For a distance $d=5d_5$~kpc and a
projected velocity of the nebula $V_n = 1000v_n$~\kms, this
lets us derive an upper limit $v_n < 0.59d_5$.

One possible explanation for this unexpectedly low velocity is that the pulsar
is moving at $V_p \sim 1500$~\kms~$\approx 2.5v_n$, but its associated
nebula, sensitive to its surroundings, does not track the ballistic
motion of the pulsar on a one-to-one basis.  The distance, $r_w$,
between the pulsar and the head of the bow-shock is determined by
pressure balance between the pulsar wind and the ram pressure produced
by the pulsar's motion, such that (ref.\ 6, 13\nocite{fk91,vm88b}):
\begin{equation}
\eta\dot{E}/4\pi r_w^2 c = \frac{4}{9}\rho V_p^2,
\label{eqn_ram}
\end{equation}
where $\rho$ is the mass
density of the ambient medium and $\eta$ is the fraction of the
pulsar's spin-down luminosity, $\dot{E}\equiv -4\pi^2 I\nu\dot{\nu}
= 2.6\times10^{36}$~erg~s$^{-1}$,
which goes into powering its relativistic wind ($\eta < 1$).  Using
the pulsar's position as measured in late 1999 (see legend to
Fig~\ref{fig_duck}), we find that $r_w = 0.036d_5$~pc
for the second epoch.
Assuming that $v_p = 2.5v_n$ (where $V_p =
1000v_p$~\kms), we can then infer that $r_w$ must have decreased by at
least 20\% between Epochs 1 and 2. A decrease in $\eta \dot{E}$ by a
similar factor can produce this effect.  However, comparison of pulsar
timing parameters from the time of Epoch 2 (A.~G.~Lyne, private
communication) with those published before Epoch 1 (ref.\ 5\nocite{mkj+91}) 
shows no significant change in $\dot{E}$, while a change in $\eta$ seems
unlikely given that changes in the efficiency of pulsar winds, if they
occur at all, most likely occur on time-scales of
$\sim10^4-10^6$~yr\cite{fs97,tab+94}.

A reasonable alternative is that the pulsar is propagating into a
density gradient so that $\rho$ is increasing as a function of time,
causing a decrease in $r_w$. However, an increasing density should
result in a change in the morphology or surface brightness of the
leading edge of \duck. In fact, the structure and brightness
of the head of \duck\ is
remarkably similar between the two epochs, and also in archival 6~cm
VLA data taken at various epochs between 1985 and 1999.  There is thus
no evidence from the morphology of \duck\ for a change in ambient
conditions between the two epochs.

It thus seems most likely that $r_w$ has not changed between the
two epochs, implying that $V_p = V_n \ll 1500$~\kms. In this case,
demanding that $t_p \approx \tau$ would mean that the pulsar had only
travelled $7'$ over its lifetime, inconsistent with its separation
of $16.1'-20.6'$ from the apparent centre of the SNR\cite{fkw94}
and thus arguing against a physical association between the two
objects.  While such chance coincidence along the line-of-sight is
not impossible\cite{gj95c}, the strong morphological evidence for an
association\cite{mkj+91,fk91,sfs89,fkw94,ckk+87} leads us to conclude that
$t_p > 39-50$~kyr~$\gg \tau$. This result depends only on assuming that
$V_p = V_n$, and that the pulsar was born near the SNR's apparent centre.

Another approach is to let the age of the system, the
ambient density and the transverse velocity of the pulsar all be
free parameters. We can then solve for these unknowns using the three
characteristic length scales of the system, namely the radius of the SNR
($R_s = 25d_5$~pc),
the stand-off distance of the bow-shock ($r_w = 0.036d_5$~pc)
and the offset of the pulsar
from the SNR's centre ($R_p = [23.4-30.0]d_5$~pc).
To determine how the unknown quantities relate to $R_s$,
we first need to determine the SNR's evolutionary state.  
Using our upper limit $v_p < 0.59d_5$
and the value measured for $r_w$,
Equation~(\ref{eqn_ram}) implies a
particle density $n_1>n_0 = 0.23\eta d_5^4$~cm$^{-3}$.  Assuming that
the SNR is expanding into a similar environment, it will be in the
radiative (or ``snowplough'') phase of its evolution\cite{sfs89,cmb88}
if $t_p > t_R = 13E_{51}^{3/14}n_1^{-4/7}$~kyr, where $E_{51}
10^{51}$~erg is the energy in the supernova explosion and assuming solar
ISM abundances. The lower limits we have already computed on $n_1$ and
$t_p$ ensure that the SNR is indeed in the radiative
phase; the radius of the SNR is then given by:\cite{sfs89,cmb88}
\begin{equation}
R_s = 7.0E_{51}^{0.221}n_1^{-0.257}t_3^{0.3}~{\rm pc},
\label{eqn_1}
\end{equation}
where $t_p = t_3$~kyr.  Meanwhile, ballistic motion
of the pulsar away from the SNR's centre gives:
\begin{equation}
R_p = 1.02 v_p t_3~{\rm pc},
\label{eqn_2}
\end{equation}
while we can rewrite Equation~(\ref{eqn_ram}) to show that:
\begin{equation}
r_w = 0.010\eta^{1/2}n_1^{-1/2}v_p^{-1}~{\rm pc}.
\label{eqn_3}
\end{equation}
Equations (\ref{eqn_1}), (\ref{eqn_2}) and (\ref{eqn_3})
can be solved simultaneously
to give $v_p = (0.17-0.25)E_{51}^{-1.04}\eta^{1.21}d_5^{2.3}$,
$n_1 = (1.3-2.6)E_{51}^{2.08}\eta^{-1.42}d_5^{-6.6}$ and $t_3 =
(93-170)E_{51}^{1.04}\eta^{-1.21}d_5^{-2.3}$.  Self-consistency
with the requirement that $n_1>n_0$ then implies $d_5 <
1.3E_{51}^{2.08}\eta^{-2.42}$, which combined with\cite{fkw94} $d_5 >
0.9$ (from \HI\ absorption) and $d_5 \approx 0.9$ (from the pulsar's
dispersion measure) justifies the choices $d_5 = 1$ and $E_{51} = 1$.

Thus two different approaches --- using either the limits on proper
motion, or the morphology of the system --- both imply a much older
and more slowly moving pulsar than results\cite{fk91} when one assumes
$t_p=\tau$. We note that a similar age and velocity to those we have just
estimated were derived before the
characteristic age of the pulsar was known, by assuming $n_1=1$ and
then solving together the equivalents of Equations (\ref{eqn_1}) and
(\ref{eqn_2}) (ref.  7\nocite{sfs89}).

For a pulsar of constant magnetic dipole moment and moment of inertia,
it can be shown that\cite{mt77}:
\begin{equation}
\frac{t_p}{\tau} = \frac{2}{n-1} \left[ 1 - \left( 
\frac{\nu}{\nu_0} \right)^{n-1} \right],
\label{eqn_age}
\end{equation}
where $\nu_0$ is the pulsar's initial rotational frequency and $n$ is
the ``braking index'', defined by $\dot{\nu} = -K\nu^n$ (where
$K$ is a constant), and expected
to be $n=3$ for a dipole rotating {\em in vacuo}.  Our lower limit on
$t_p/\tau$ from proper motion alone implies that $n<1.83$, while the
value derived from the morphology of the system argues that $n<1.33$.
These values are significantly less than those derived for young
pulsars\cite{lps88,kms+94}, but are similar to $n=1.4\pm0.2$ observed
for the Vela pulsar\cite{lpgc96}, a pulsar which also resembles
PSR~B1757--24 in its value of $\tau$ and its tendency to
glitch.\cite{lkb+96} One alternative possibility is that the magnetic
field of PSR~B1757--24 is growing as a function of
time\cite{bah83,br88}.  In this case Equation~(\ref{eqn_age}) does not
hold and $K$ is no longer a constant; $\tau$ is then
an approximate time-scale for field growth.

Further data can support the conclusion that PSR~B1757--24 is much
older than $\tau = \nu/2\dot{\nu}$: multi-epoch VLBI observations are
expected to show westward motion of the pulsar at the level of
$\sim$10~milliarcsec~yr$^{-1}$, X-ray observations of SNR~G5.4--1.2
should show emission characteristic of an old rather than young SNR,
and a braking index for PSR~B1757--24 measured from timing data would
be expected to lie in the range predicted above.  The low braking index
of the Vela pulsar\cite{lpgc96}, discrepant pulsar/SNR ages in other
pulsar/SNR assocations\cite{br88,shmc83}, and the fact that many pulsars
with small characteristic ages have no associated SNR\cite{bgl89,sgj99},
all suggest that PSR~B1757--24 is not a statistical anomaly. If other
pulsars are indeed older than they seem, our understanding of pulsar
velocities, asymmetries in supernova explosions, the fraction of
supernovae that produce pulsars and the physics of neutron star
structure and cooling must all be reconsidered\cite{cc98,fgw94,bj98,usnt93}.

\bibliographystyle{nature}
\bibliography{journals,modrefs,psrrefs}

\begin{acknowledge}
We thank Vicky Kaspi and Deepto Chakrabarty for useful discussions,
Namir Kassim for supplying 90~cm data on G5.4--1.2 and Andrew Lyne for
providing timing data on PSR~B1757--24.  The National Radio Astronomy
Observatory is a facility of the National Science Foundation operated
under cooperative agreement by Associated Universities, Inc.  B.M.G.
acknowledges the support of NASA through a Hubble Fellowship awarded by
the Space Telescope Science Institute.
\end{acknowledge}

{\raggedright Correspondence should be addressed to B.M.G. 
(e-mail:
bmg@space.mit.edu)}.


\begin{table}[hbt]
\caption{Very Large Array observations of the
pulsar-powered nebula \duck\ used for proper motion measurements.  
Both observations
used the same pointing centre, namely (J2000) Right Ascension
(RA)~18$^{\rm h}$01$^{\rm m}$00.035$^{\rm s}$, Declination (Dec)
--24$^{\circ}51'27.260''$.  Absolute flux densities were determined
using observations of the source 3C~286, while antenna gains were
calibrated using observations every 15--20~min of the extragalactic
source PMN~J1751--2524, displaced $2\hbox{$.\!\!^\circ$}1$ from \duck.
The resulting images were formed using natural weighting and multi-frequency
synthesis with 100~mas pixels, deconvolved using a maximum entropy
algorithm, and then smoothed with a circular Gaussian of FWHM
$0\hbox{$.\!\!^{\prime\prime}$}90$.}
\vspace{1cm}
\begin{center}
\begin{tabular}{lcc} \hline
              & Epoch 1  & Epoch 2 \\
Date Observed & 1993 Feb.~02 & 1999 Oct.~23  \\
Array Configuration & BnA & BnA \\
On-source integration time (h) & 3.3 & 2.8 \\
Centre Frequency (GHz) &  8.44 & 8.46 \\
Bandwidth (MHz)   & 75 & 100 \\ \hline
\end{tabular}
\end{center}
\label{tab_obs}
\end{table}

\clearpage

\clearpage

\end{document}